\documentclass[fp,twocolumn]{jpsj3}

\usepackage{txfonts}

\usepackage{graphicx}
\usepackage{dcolumn}
\usepackage{bm}
\usepackage{color}
\usepackage{mathrsfs}

\title{ 
Phonon-Induced Current Noise in Single-Walled Carbon Nanotubes across the Ballistic-Diffusive Crossover}
\author{Aina Sumiyoshi$^1$ and Takahiro Yamamoto$^{1,2}$}

\inst{$^1$ Department of Physics, Tokyo University of Science, Kagurazaka 1-3 Shinjuku, Tokyo 162-8601, Japan \\
$^2$ RIST, Tokyo University of Science, Kagurazaka 1-3 Shinjuku, Tokyo 162-8601, Japan} 

\abst{
We theoretically elucidate the system length ($L$) dependence of phonon-induced current noise in carbon nanotubes at room temperature over a broad 
range, encompassing the quantum ballistic and classical diffusive regimes. The power spectral density for the current noise is maximally enhanced 
when $L$ is comparable to the mean free path $L_0$ of an electron. In the ballistic limit of $L/L_0\ll 1$, the power spectral density increases in proportion to $L$, 
whereas in the diffusive limit of $L/L_0\gg 1$, it shows a power-law decay $L^{-\alpha}$ with a scaling parameter $\alpha=3.81$. The noise decay for 
single-walled carbon nanotubes is faster than that previously predicted based on a simple model because of the various electron--phonon scattering processes 
and the complex energy dependence of the phonon relaxation time.
}

\begin{document}
\maketitle

\section{Introduction~\label{sec:1}}                                                           
                           
With the remarkable advances in nanotechnology since the late 20th century, quantum transport in mesoscopic systems has been extensively studied, 
both theoretically and experimentally~\cite{rr:datta,rr:imry}. In these systems, quantum effects become prominent. Various quantum phenomena, such as 
quantized conductance~\cite{rr:Wees}, universal conductance fluctuation~\cite{rr:lee_stone}, and the Aharonov-Bohm effect~\cite{rr:tonomura,rr:Washburn}, 
have been observed. These phenomena are all related to the average current; however, it has been recently revealed that the current noise, 
which is enhanced in small systems, also contains a large amount of physical information~\cite{rr:Jong,rr:Blanter,rr:Martin,rr:Piatrusha,rr:kobayashi}. One
example is the fractional charge in the fractional quantum Hall effect discovered via current noise measurements~\cite{rr:Saminadayar,rr:Picciotto}. 
Consequently, understanding and controlling current noise has become increasingly important.

Current noise can generally be categorized into {\it external} and {\it internal} noise. External noise  includes $1/f$ noise (also called flicker noise) and random 
telegraph noise, which appear to originate from defects and impurity distributions within the conductor and its surroundings. It also includes 
contributions from external circuits, such as power supplies and resistors connected to the conductor. While these types of noise are critical issues in the 
design of electronic devices, condensed matter physics focuses on internal noise, which exists even in pristine conductors connected to ideal circuits without 
external interference. Among the various types of internal noise, {\it phonon-induced} current noise always exists at finite temperatures and cannot be eliminated; 
consequently, it has attracted much experimental and theoretical attention. Nevertheless, as shown below, our understanding of current noise arising from {\it dynamic} 
scattering such as electron--phonon interactions remains incomplete, especially in comparison with {\it static} scattering mechanisms like impurity scattering.

In a mesoscopic conductor sandwiched between two symmetric leads, the intensity of the current noise at low temperatures (i.e., the quantum shot noise $S$)
is expressed as
\begin{equation}
S\propto \sum_{n=1}^M \zeta_n(1-\zeta_n)
\label{eq:landaure_S}
\end{equation}
using the transmission function $\zeta_n$ at the Fermi energy~\cite{rr:Lesovik,rr:buttiker00}. Here, $n$ labels each conduction channel in the lead
and $M$ is the total number of conduction channels. However, Eq.~(\ref{eq:landaure_S}) does not hold in the presence of electron--phonon scattering, where energy 
conservation for electrons is violated and the concept of transmission probability $\zeta$ itself becomes ambiguous. In fact, as the system length $L$ for 
a conductor increases to more than the mean free path $L_0$, electron--phonon scattering becomes dominant for electronic transport, leading to 
a crossover from quantum ballistic to classical diffusive transport. In the diffusive transport regime, several theoretical studies have investigated the 
$L$-dependence of current noise in one-dimensional (1D) systems~\cite{rr:Shimizu,rr:Beenakker,rr:Nagaev1992,rr:Nagaev1995,rr:Naveh}; however,
no clear consensus has been reached, as the results strongly depend on the theoretical treatment of electron--phonon interactions. Let alone, 
no large-scale atomistic quantum transport studies have been conducted for realistic materials such as carbon nanotubes (CNTs)~\cite{rr:iijima, rr:saito, rr:hamada}. 
In addition, the $L$ dependence of current noise in the crossover region  ($L/L_0\approx 1$) between the quantum ballistic and classical diffusive transport regimes 
remains largely unexplored in realistic materials and is still an open question. 

In this study, we aim to clarify how phonon-induced scattering affects current noise during the crossover from ballistic to diffusive transport in realistic 
1D systems, particularly CNTs. To this end, we develop a large-scale atomistic quantum transport simulation framework that incorporates 
electron--phonon interactions. Using this approach, we investigate the $L$ dependence of the zero-frequency current noise $S$
not only in the ballistic and diffusive regimes, but also in the crossover region around $L/L_0=1$. This study addresses a critical gap in our understanding 
of nonequilibrium current noise phenomena in realistic low-dimensional materials.

\section{Theory and Method}
It is generally impossible to exactly solve many-body quantum transport problems that involve the electron-nucleus interactions of interest here and thus 
appropriate approximations are required. In this study, we use the Born-Oppenheimer (BO) approximation, in which the total wave function is decomposed 
into {\it nuclear motion} and {\it electronic states that depend on the nuclear positions}. The BO approximation is known to remain valid even for current-carrying 
nonequilibrium systems, provided that the time scales of nuclear motion and electronic dynamics differ significantly. 

Furthermore, we assume that nuclear quantum effects are negligible, so that the motion of the nuclei can be treated using classical equations of motion on an 
adiabatic potential energy surface. We assume that the motion of the nuclei immediately relaxes to a thermal equilibrium state and determine the motion 
of the nuclei at finite temperatures using classical molecular dynamics (MD) simulation within the NTV ensemble with the velocity scaling scheme. We also adopt 
the Tersoff potential as a reliable force field for carbon-carbon interaction in a single-walled CNT (SWCNT)~\cite{rr:lindsay}. 

The theoretical scheme and simulation method for electronic states and current noise are explained in the following subsection.

\subsection{Basics of current noise}
Let us consider instantaneous, current-carrying steady states in which a time-dependent current $J(t)$ flows through a conductor. The average current is 
given by
\begin{eqnarray}
\langle J\rangle=\frac{1}{N}\sum_{k=1}^N \overline{J_k},\quad
\overline{J_k}=\lim_{\tau\to\infty}\frac{1}{\tau}\int_{-\tau/2}^{\tau/2} \!\!\! dt~J_k(t),
\end{eqnarray}
where $\overline{J_k}$ is the average current over the long time interval $\tau$ for a trial $k(=1,2,\cdots,N)$ and $\langle J\rangle$ is the ensemble 
average of these long-time-averaged currents for $N$ trials. We then define the current deviation as
\begin{equation}
\Delta J_k(t) = J_k(t) - \langle J\rangle.
\end{equation}
A standard way to quantify these deviations is via their mean-square value, 
\begin{eqnarray}
\langle \Delta J^2\rangle= \frac{1}{N}\sum_{k=1}^N {\overline{(\Delta J_k)^2}}
\end{eqnarray}
with
\begin{eqnarray}
{\overline{(\Delta J_k)^2}}=\lim_{\tau\to\infty}\frac{1}{\tau}\int_{-\tau/2}^{\tau/2} \!\!\! dt~(\Delta J_k(t))^2.
\end{eqnarray}
By performing a Fourier transform from the time domain ($t$) to the angular frequency domain ($\omega$): 
\begin{equation}
\Delta J_k(t) = \frac{1}{2\pi}\int_{-\infty}^{\infty}\!\!\! d\omega~ \Delta J_k(\omega) e^{-i\omega t},
\end{equation}
the mean-square value is rewritten as
\begin{equation}
\left\langle \Delta J^2\right\rangle=\frac{1}{2\pi}\int_0^\infty \!\!\! d\omega~S(\omega),
\end{equation}
where $S(\omega)$ is the power spectral density (PSD), which is defined by
\begin{eqnarray}
S(\omega)&=&\lim_{\tau\to\infty}\frac{2}{\tau}\langle\lvert\Delta J(\omega)\rvert^2\rangle\\
&=&\lim_{\tau\to\infty}\frac{2}{\tau}\int_{-\tau/2}^{\tau/2} \!\!\!\!\!\! dt\int_{-\tau/2}^{\tau/2}\!\!\!\!\!\!dt' \langle\Delta J(t)\Delta J(t')\rangle
e^{i\omega(t-t')}.
\label{eq:definition_PSD}
\end{eqnarray}
Note that $\Delta J_k(-\omega)=\Delta J_k(\omega)^*$ and then $S(-\omega)=S(\omega)$.

Our focus here is on the {\it white noise} regime at low frequencies, where the PSD does not depend on $\omega$. Note that we neglect $1/f$ noise 
at low frequencies in the present study. Thus, the zero-frequency limit,
\begin{equation}
S(0)=\lim_{\rm \tau\to\infty}\frac{2\langle(\Delta Q(\tau))^2\rangle}{\tau},
\label{eq:white_noise_S0}
\end{equation}
can be used as the PSD in the white noise region.
The quantity $\Delta Q_k(t):=Q_k(t)-\langle Q\rangle$ represents the fluctuation in the cumulative charge $Q_k(t)$ flowing through the conductor 
over time interval $\tau$, where $Q_k(\tau)$ is defined by
\begin{equation}
Q_k(\tau) = \int_{0}^{\tau}\!\!\! dt~ J_k(t),
\label{eq:Q_tau}
\end{equation}
and thus Eq.~(\ref{eq:white_noise_S0}) links the time-domain view of charge transport noise to its frequency-domain representation. 
For convenience, in Eq.~(\ref{eq:Q_tau}), the interval of integration has been shifted from $[-\tau/2, \tau/2]$ to $[0, \tau]$.

\subsection{Simulation model and method}

\begin{figure}
\begin{center}
\includegraphics[width=90mm]{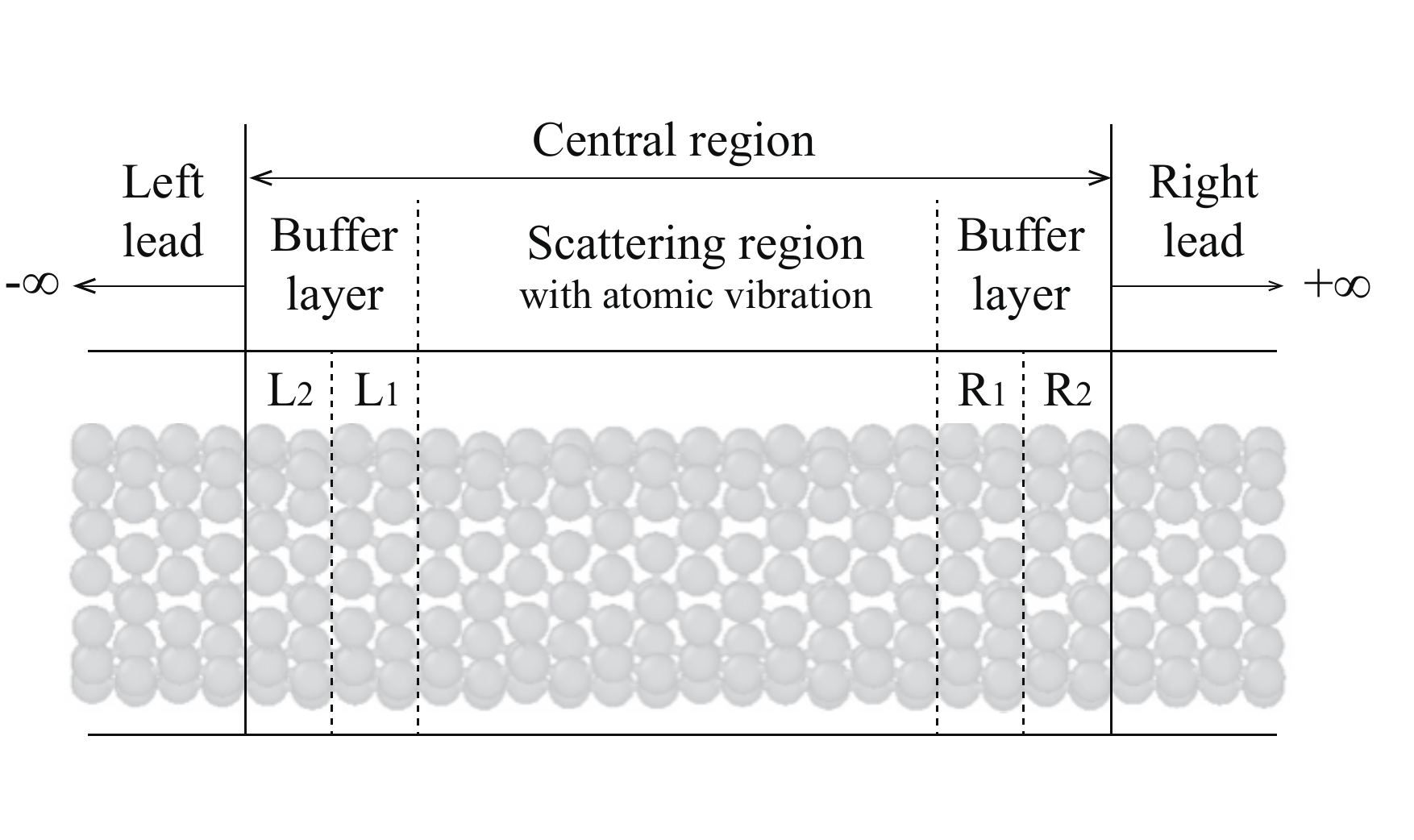}
\caption{(Color online) Simulation setup of Open-TDSE+MD method consisting of three regions: central region with finite length and semi-infinite left and 
right leads without atomic vibration. The central region is composed of a scattering region with atomic vibration and two buffer layers without vibration, 
each of which is composed of two unit cells (L$_1$ and L$_2$ or R$_1$ and R$_2$). }
\label{fig:simulation_model}
\end{center}
\end{figure}

Our research group recently developed a powerful quantum transport simulation method called OpenTDSE+MD. This method can be used to calculate 
the time-dependent current in a conductor sandwiched between two semi-infinite leads~\cite{rr:ishizeki2017,rr:ishizeki2018,rr:ishizeki2020}. 
In this section, we give a brief overview of this method.

Figure~\ref{fig:simulation_model} shows the simulation setup of the Open-TDSE+MD method, which consists of three regions: the central region, the left lead, 
and the right lead. The central region is composed of a scattering region that contains a conductor with atomic vibration and two buffer layers, each containing 
two unit cells without atomic vibration. The buffer layers are set as the locations where electrical current is calculated, as described later. The two ends of the central 
region are respectively connected to the semi-infinite leads without atomic vibration. Thus, the Hamiltonian operator ${\hat{\mathscr H}}(t)$ for the whole system 
is expressed as
\begin{eqnarray}
{\hat{\mathscr H}}(t)={\hat{\mathscr H}}^{\rm LL}+{\hat {\mathscr H}}^{\rm LC}+{\hat {\mathscr H}}^{\rm CC}(t)+{\hat {\mathscr H}}^{\rm CR}+{\hat {\mathscr H}}^{\rm RR},
\end{eqnarray}
where ${\hat {\mathscr H}}^{\rm CC}(t)$ is the time-dependent Hamiltonian in the central region, ${\hat {\mathscr H}}^{\rm LL(RR)}$ is the time-independent 
Hamiltonian in the left (right) lead, and ${\hat {\mathscr H}}^{\rm LC(CR)}$ is a time-independent Hamiltonian that connects the central region 
and the left (right) lead.

For the material of interest in this work, namely SWCNTs, the $\pi$-orbital tight-binding Hamiltonian is adopted and ${\hat {\mathscr H}}^{\rm CC}(t)$ is expressed as
\begin{eqnarray}
{\hat {\mathscr H}}^{\rm CC}(t)=\sum_{\langle i,j \rangle}\gamma_{ij} (t){\hat c}_i^\dagger {\hat c}_j,
\end{eqnarray}
where the origin of energy ($\varepsilon=0$~eV) is chosen to coincide with the on-site energy of a carbon atom, and the summation runs over 
nearest-neighbor pairs $\langle i,j\rangle$. The operators $\hat c_i^{\dagger}$ and $\hat c_i$ are fermionic creation and annihilation operators, respectively, 
for an electron in the $\pi$ orbital of $i$th atom. The time-dependent hopping integral between atoms $i$ and $j$ is evaluated with Harrison's rule~\cite{rr:harrison}:
\begin{eqnarray}
\gamma_{ij}(t)=\gamma_0\frac{|{\bm R}_{0,i}-{\bm R}_{0,j}|^2}{|{\bm R}_i(t)-{\bm R}_j(t)|^2},
\end{eqnarray}
where $\gamma_0=-2.7$eV, ${\bm R}_{0,i}$ is the equilibrium position of the $i$th carbon atom and ${\bm R}_i(t)$ is the position of the $i$th atom at time $t$, 
which is calculated by the MD simulation. In this method, the electron--phonon interaction is incorporated through the time dependency in ${\hat {\mathscr H}}^{\rm CC}(t)$.
The atoms in the two buffer layers are fixed at the equilibrium positions and thus the matrix elements of 
${\hat {\mathscr H}}^{\rm LC(CR)}$ and those of ${\hat {\mathscr H}}^{\rm CC}$ in the buffer layers  are independent of time. In this study, the Hamiltonian of 
the left (right) lead is given by that for a pristine SWCNT without any disorder as 
${\hat {\mathscr H}}^{\rm LL(RR)}=\gamma_0\sum_{\langle i,j \rangle}{\hat c}_i^\dagger {\hat c}_j$. The current noise vanishes at $T=0$ 
for this setup.

Let us consider the current-carrying wave function $\bm{\Psi}^{\rm C}_{\pm,\varepsilon,n}(t)$ in the central region for an incident electron with energy 
$\varepsilon$ from the $n$th channel in a lead. The subscript $+$ ($-$) labels the right-going (left-going) state of an electron injected from the left (right) lead. 
To obtain  $\bm{\Psi}^{\rm C}_{\pm,\varepsilon,n}(t)$, we solve the time-dependent Schr{\" o}dinger equation (TDSE) for an open system:
\begin{eqnarray}
i\hbar \frac{\partial \bm{\Psi}^{\rm C}_{\pm,\varepsilon,n}(t)}{\partial t} 
= {\mathcal H}^{\rm{CC}}(t)\bm{\Psi}^{\rm C}_{\pm,\varepsilon,n}(t) + \bm{S}_{\pm,\varepsilon,n}(t) + \bm{D}_{\pm,\varepsilon,n}(t),
\label{eq:open_TDSE}
\end{eqnarray}
where ${\mathcal H}^{\rm{CC}}(t)$ is the Hamiltonian matrix for the central region, $\bm{S}_{\alpha,\varepsilon,n}(t)$ is the influx of electrons from 
lead L/R for $\alpha=+/-$, which is given by
\begin{eqnarray}
 \bm{S}_{\pm,\varepsilon,n}(t) = {\mathcal H}^{\rm{CL/CR}}\frac{\boldsymbol{\phi}_{\pm,\varepsilon,n}}{\sqrt{v_n}}\exp\left(-\frac{i\varepsilon t}{\hbar}\right).
 \end{eqnarray}
 Here, $\bm{\phi}_{\pm,\varepsilon,n}$ is the wave function vector in the isolated left ($+$) and right ($-$) leads normalized by the number of atoms in 
 a unit cell and $v_n$ is the group velocity of $\bm{\phi}_{\pm,\varepsilon_{\rm{F}},n}$ toward the central region. $\bm{D}_{\pm,\varepsilon,n}$ is the 
 outflux of electrons from the central region into the two leads, which is given by
 \begin{eqnarray}
\bm{D}_{\pm,\varepsilon,n}(t) = \big(\Sigma^{\rm{L}}(\varepsilon) + \Sigma^{\rm{R}}(\varepsilon)\big) \bm{\Psi}^{\rm{C}}_{\pm,\varepsilon,n}(t)
 \label{eq:outflux}
 \end{eqnarray}
using the retarded self-energy matrix $\Sigma^{\rm L/R}(\varepsilon)$, due to the left/right lead within the wide-band limit scheme. Details of the derivation of 
Eqs.~(\ref{eq:open_TDSE})--(\ref{eq:outflux}) and their numerical treatment can be found elsewhere~\cite{rr:ishizeki2017}.

Once we obtain ${\bm \Psi}^{\rm{C}}_{\pm,\varepsilon,n}(t)$, we can calculate the {\it rightward} current $j_{\varepsilon,n}^{+}(t)$ using 
\begin{equation}
j_{\varepsilon,n}^{+}(t)=
\frac{2a}{\hbar}{\rm Im}\big\lbrack\big(\bm{\Psi}^{\rm{R}_2}_{+,\varepsilon,n}(t)\big)^{\dagger}{\mathcal H}^{\rm{R}_2\rm{R}_1}(t)\bm{\Psi}^{\rm{R}_1}_{+,\varepsilon,n}(t)\big\rbrack,
\end{equation}
and the {\it leftward} current $j_{\varepsilon,n}^{-}(t)$ using
\begin{equation}
j_{\varepsilon,n}^{-}(t)=
\frac{2a}{\hbar}{\rm Im}\big\lbrack\big(\bm{\Psi}^{\rm{L}_2}_{-,\varepsilon,n}(t)\big)^{\dagger}{\mathcal H}^{\rm{L}_2\rm{L}_1}(t)\bm{\Psi}^{\rm{L}_1}_{-,\varepsilon,n}(t)\big\rbrack,
\end{equation}
where $a$ is the length of a unit cell, $\bm{\Psi}^{\rm{R}_{1(2)}}_{+,\varepsilon,n}(t)$ and $\bm{\Psi}^{\rm{L}_{1(2)}}_{-,\varepsilon,n}(t)$ are respectively 
the wave function vectors in buffer layers $\rm{R}_{1(2)}$ and $\rm{L}_{1(2)}$, and ${\mathcal H}^{\rm{R}_1\rm{R}_2}$ (${\mathcal H}^{\rm{L}_1\rm{L}_2}$) is 
the Hamiltonian matrix connecting $\rm{R}_{1}$ and $\rm{R}_2$ ($\rm{L}_{1}$ and $\rm{L}_2$). The current $j_{\varepsilon,n}^{\pm}(t)$ can be regarded 
as the time-dependent transmission probability of an electron with energy $\varepsilon$ shot from channel $n$  in the left/right lead. Thus, according to 
the Landauer formula, the electrical current can be expressed as
\begin{equation}
J^{\pm}(t)=J_0\sum_{n}j_{\varepsilon_{\rm F},n}^{\pm}(t),\quad J_0:=\frac{2e^2}{h}V
\label{eq:dimless_current0}
\end{equation}
in terms of the dimensionless current at the Fermi energy $\varepsilon_{\rm F}$ when both the bias voltage $V$ between the two leads and the electron temperature 
$T_{\rm e}$ are very small. In this study, we assume that the electron temperature is $T_{\rm e}=0$~K and that the phonon temperature is finite (300~K), which is 
a reasonable assumption for metallic SWCNTs.

In Eq.~(\ref{eq:dimless_current0}), $e(<0)$ is the electron charge, $J_0$ is the electrical current for a single channel in the case of perfect transmission, 
$j_{\varepsilon_{\rm F},n}^{\pm}(t)=1$, and $J^{\pm}(t)$ is related to $J(t)$ introduced in the previous section via $J(t):=J^+(t)=-J^{-}(t)$.

\section{Application to Metallic Single-Walled Carbon Nanotubes}
\subsection{Time-dependent current with noise due to electron--phonon scattering}
In this section, we show the phonon-induced current noise in metallic (5,5) SWCNTs with various lengths obtained using the Open-TDSE+MD method. 
The Fermi energy is taken as the charge neutral point (CNP), $\varepsilon_{\rm F}=0$~eV. Thus, the summation index $n$ in Eq.~(\ref{eq:dimless_current0})
runs over the two CNPs. Hereafter, we refer to these two points as the K and K$'$ points, in analogy with the Dirac points in graphene, to which they correspond in the unrolled Brillouin zone.
Thus, the dimensionless current is given by
\begin{equation}
j(t):=\frac{J(t)}{|J_0|}=j_{\varepsilon_{\rm F},{\rm K}}^{+}(t)+j_{\varepsilon_{\rm F},{\rm K'}}^{+}(t)
\label{eq:dimless_current}
\end{equation}
with $\varepsilon_{\rm F}=0$~eV. Hereafter, the superscript $+$ and the subscript $\varepsilon_{\rm F}$ are omitted from $j_{\varepsilon_{\rm F},{\rm K}}^{+}(t)$, 
which is expressed simply as $j_{\rm K}(t)$. In the present study, we set $t=0$ at a certain time in the steady state (i.e., after sufficient time has passed since 
the current began to flow in the central region). We use $\Delta t=0.1$~fs as the time step.

\begin{figure}
\begin{center}
\includegraphics[width=90mm]{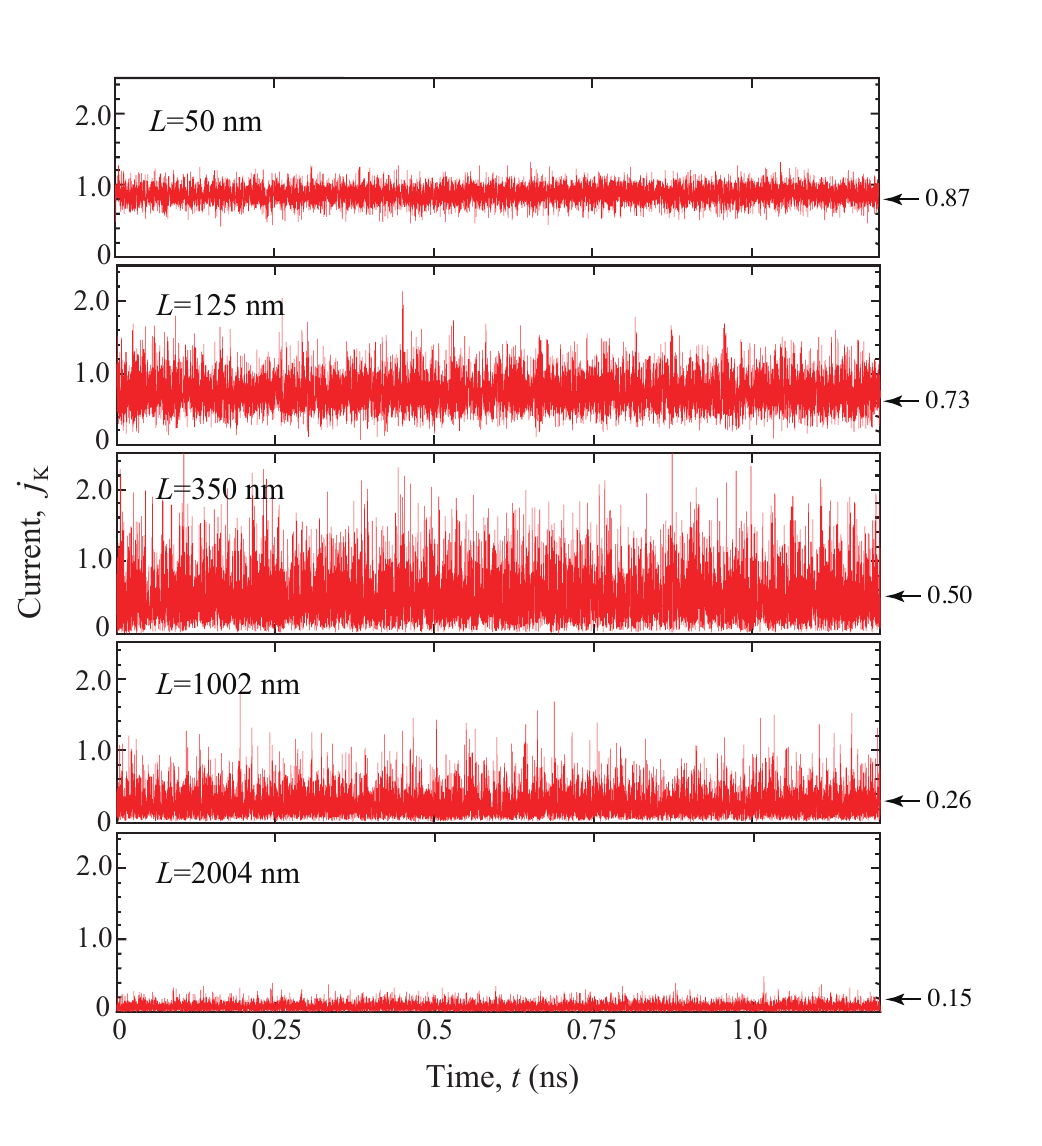}
\caption{(Color online) Time dependence of current carried by electrons at the K point in (5,5) SWCNTs with lengths $L=50$, $125$, $350$, $1002$, and $2004$~nm 
at $300$~K, where current is scaled by $J_0=2e^2V/h$. The average current for each length $L$ is indicated by the arrow.}
\label{fig:td_current}
\end{center}
\end{figure}

Figure~\ref{fig:td_current} shows $j_{{\rm K}}(t)$ flowing through (5,5) SWCNTs with lengths of scattering region $L=50$, $125$, $350$, $1002$, and $2004$~nm at room 
temperature ($T$=300~K). Note that the length $L$ is given by $L=n_{\rm uc}a$, where $a=0.25$~nm is the length of unit cell in the (5,5) SWCNT 
and $n_{\rm uc}$ is the number of unit cells in the scattering region.
As indicated by the arrows in Fig.~\ref{fig:td_current}, the average current $\langle j_{{\rm K}}(t)\rangle$ decreases monotonically with 
increasing $L$, whereas the current noise does not vary monotonically with $L$; instead, it exhibits a maximum at $L=350$~nm, which is approximately the mean 
free path ($L_0=346$~nm) as discussed later.
A detailed discussion on the $L$ dependence of $\langle j\rangle$ and $\langle \Delta j^2\rangle$ is given in the following sections.

\subsection{Length dependence of conductance}
Using Eq.~(\ref{eq:dimless_current}), the conductance $G$ in a (5,5) SWCNT is given by
\begin{equation}
G:=\lim_{V\to 0}\frac{\langle J\rangle}{V}=\frac{2e^2}{h}(\zeta_{\rm K}+\zeta_{\rm K'}),
\label{eq:landauer}
\end{equation}
where $\zeta_{\rm K(K')}:=\langle j_{\rm K(K')}\rangle$ is the transmission function for an electron injected from the left lead at the K (K$'$) point with 
the Fermi energy to the scattering region with length $L$. This is just the Landauer formula for conductance in the steady state~\cite{rr:datta,rr:imry}.

Figure~\ref{fig:LD_conductance} shows the $L$ dependence of $G$ in a (5,5) SWCNT at $T=300$~K. In the short-$L$ limit (ballistic limit), $G$ is independent 
of $L$ and its value approaches ${4e^2}/{h}(:=2G_0)$. On the other hand, in the long-$L$ limit (diffusive limit), $G$ obeys Ohm's law and is thus inversely proportional 
to $L$. Similar conductance (or resistance) behavior has been obtained using the wave packet diffusion method~\cite{rr:ishii01,rr:ishii02}. The obtained $L$ dependence 
of $G$ can be understood as follows.

The effective transmission function, including the phase breaking effect due to electron--phonon scattering, can be phenomenologically expressed as
\begin{equation}
\zeta_{\rm K(K')}=\frac{L_0}{L_0+L},
\label{eq:datta}
\end{equation}
where $L_0$ is the mean free path of electrons due to electron--phonon scattering~\cite{rr:datta}. Substituting Eq.~(\ref{eq:datta}) into Eq.~(\ref{eq:landauer}) yields the following expression for the conductance:
\begin{equation}
G=2G_0\frac{L_0}{L_0+L}.
\label{eq:landauer01}
\end{equation}
From Eq.~(\ref{eq:landauer01}), the conductance reaches $2G_0$ in the ballistic limit ($L/L_0\ll 1$), whereas it reaches $2G_0(L_0/L)$ in the diffusive limit 
($L/L_0\gg 1$). In fact, the simulation data in Fig.~\ref{fig:LD_conductance} can be well fitted by Eq.~(\ref{eq:landauer01}) with $L_0=346$~nm. The obtained $L_0$ is
in agreement with the previous theoretical reports in Refs.~[\citen{rr:suzuura, rr:ishii02,rr:ishizeki2017}]. Since the lengths of the experimentally synthesized SWCNTs are in the range of several hundred nanometers to 
several micrometers, the SWCNTs  are in the ballistic-diffusive crossover region at room temperature.

\subsection{Length dependence of current noise}
The current deviation $\Delta J(t)$ of a (5,5) SWCNT is expressed as
\begin{equation}
\Delta J(t)=\Delta J_{\rm K}(t)+\Delta J_{\rm K'}(t)
\label{eq:dimless_current_fluctuation}
\end{equation}
with $\Delta J_{\rm K(K')}:=|J_0|( j_{{\rm K(K')}}-\langle j_{{\rm K(K')}}\rangle)$, and therefore the PSD is expressed as
\begin{equation}
S(0)=S_{\rm KK}(0)+S_{\rm K'K'}(0)+S_{\rm KK'}(0)+S_{\rm K'K}(0),
\end{equation}
where $S_{\alpha\beta}$ ($\alpha,\beta=$K and K$'$) is defined by
\begin{equation}
S_{\alpha\beta}(0)=\lim_{\tau\to\infty}\frac{2}{\tau}\int_{0}^{\tau} \!\!\! dt\int_{0}^{\tau}\!\!\!  dt' 
\langle\Delta J_\alpha(t)\Delta J_\beta(t')
\rangle.
\end{equation}
Since the low-frequency current noise is affected only by the long-wavelength phonons, scattering occurs exclusively within the K and K' valleys and 
$S_{\rm KK'(K'K)}$ is negligible. In addition, the electronic states at the K and K$'$ valleys of the SWCNTs exhibit equivalent electron--phonon coupling, which results in identical PSDs, 
(i.e., $S_{\rm KK}=S_{\rm K'K'}$). Therefore, we focus on $S_{\rm KK}$ in the following.

\begin{figure}[t]
\begin{center}
\includegraphics[width=75mm]{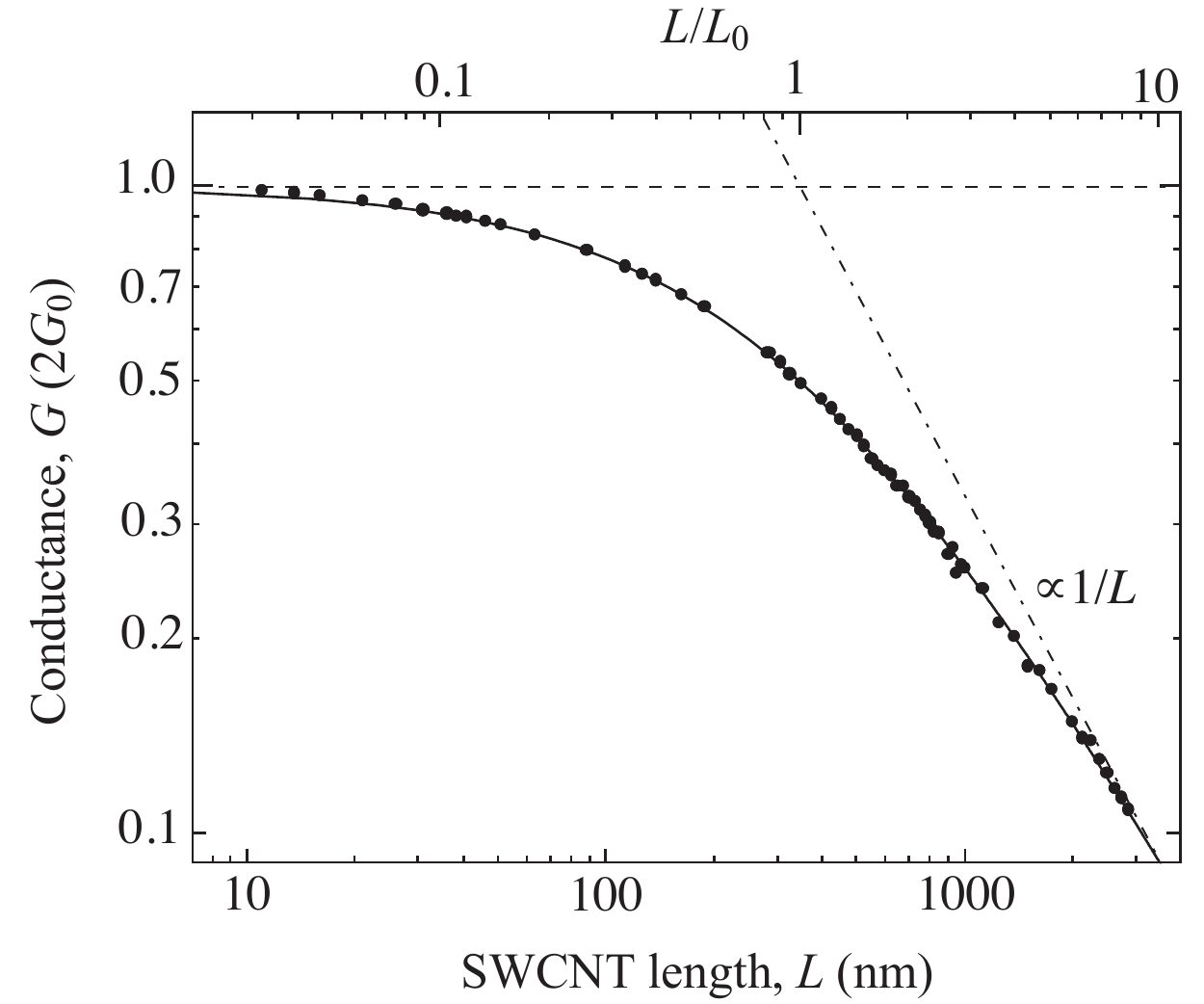}
\caption{Length dependence of conductance of (5,5) SWCNT at $300$~K. Here, $G_0=2e^2/h$ is the quantum conductance. 
The dashed and dotted-dashed lines represent the asymptotic lines for the conductance in the ballistic and diffusive limits, respectively.
The solid curve is the theoretical curve in Eq.~(\ref{eq:landauer01}) with $L_0=346~{\rm nm}$.}
\label{fig:LD_conductance}
\end{center}
\end{figure}

\begin{figure}[t]
\begin{center}
\includegraphics[width=80mm]{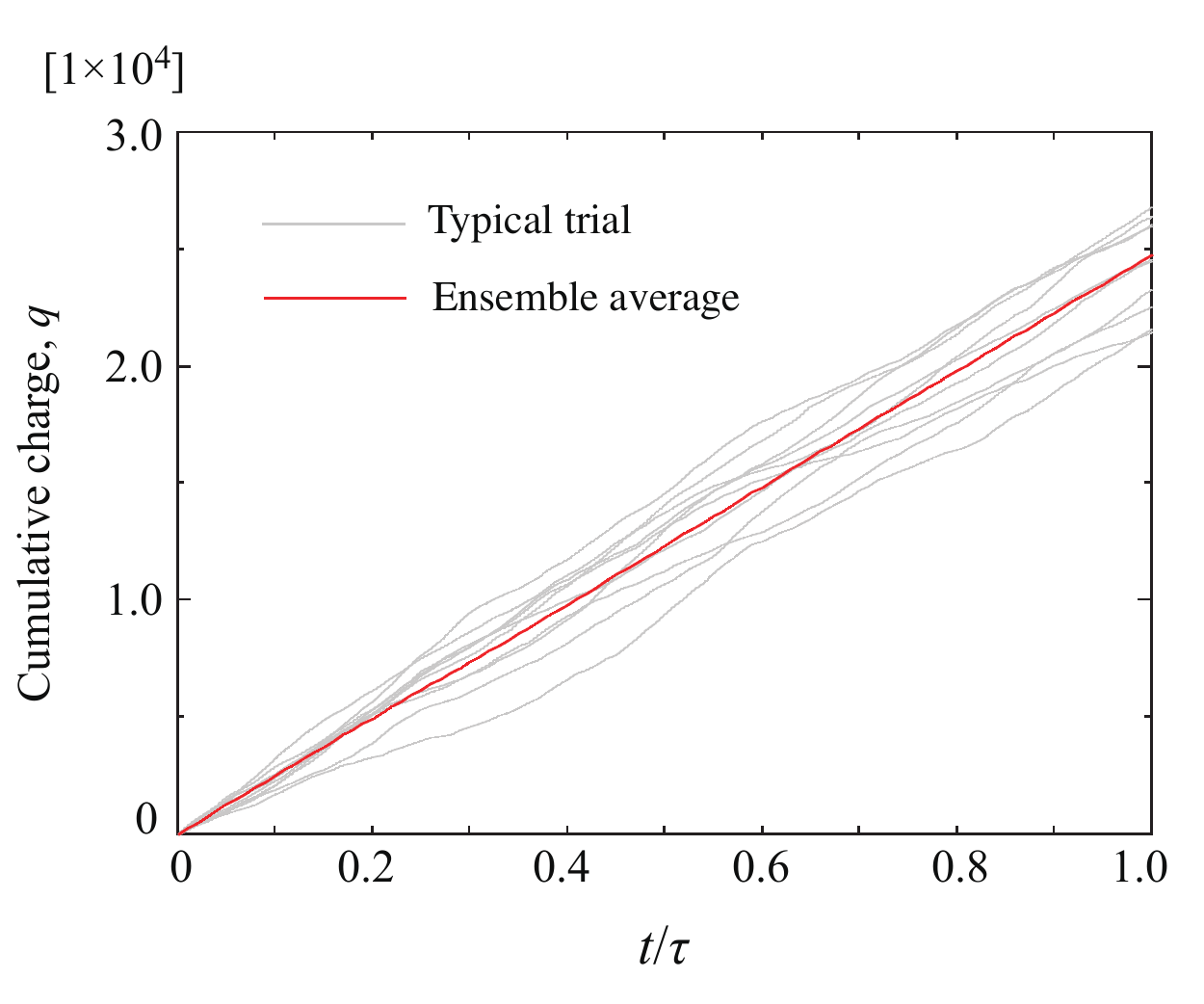}
\caption{(Color online) Time dependence of total charge (dimensionless cumulative charge) $q(t)$ passing through (5,5) SWCNT with $L=350$~nm at 300~K. Here, 
$\tau$ is the measurement time interval; it is taken to be 5.0~ps. See the main text for the definition of $q(t)$. The gray curves are the results for 10 typical trials and the red curve is the ensemble average of 240 trials.}
\label{fig:cumulative_charge}
\end{center}
\end{figure}

Figure~\ref{fig:cumulative_charge} shows the time dependence of the cumulative charge $q_k(t)=Q_k(t)/Q_0$ passing through a (5,5) SWCNT with $L= 350$~nm 
for various trials under various initial conditions at $T=300$~K, where $k$ is the label of the trials and $Q_0$ is defined as $Q_0=J_0\Delta t$. 
In Fig.~\ref{fig:cumulative_charge}, $\tau$ is the measurement time interval; it is taken to be $5.0$~ps, which is longer than the mean free time $\tau_0=0.3$~ps. 
The gray curves are the cumulative charges $q_k(t)$ for 10 typical trials and the red curve indicates the ensemble average of the 240 trials: 
\begin{equation}
\langle q(t)\rangle=\frac{1}{N}\sum_{k=1}^Nq_k(t)
\end{equation}
with $N=240$. We can see in Fig.~\ref{fig:cumulative_charge} that $\langle q(t)\rangle$ is proportional to $t$. We thus judge that the ensemble average has 
sufficient statistical accuracy.

\begin{figure}[t]
\begin{center}
\includegraphics[width=85mm]{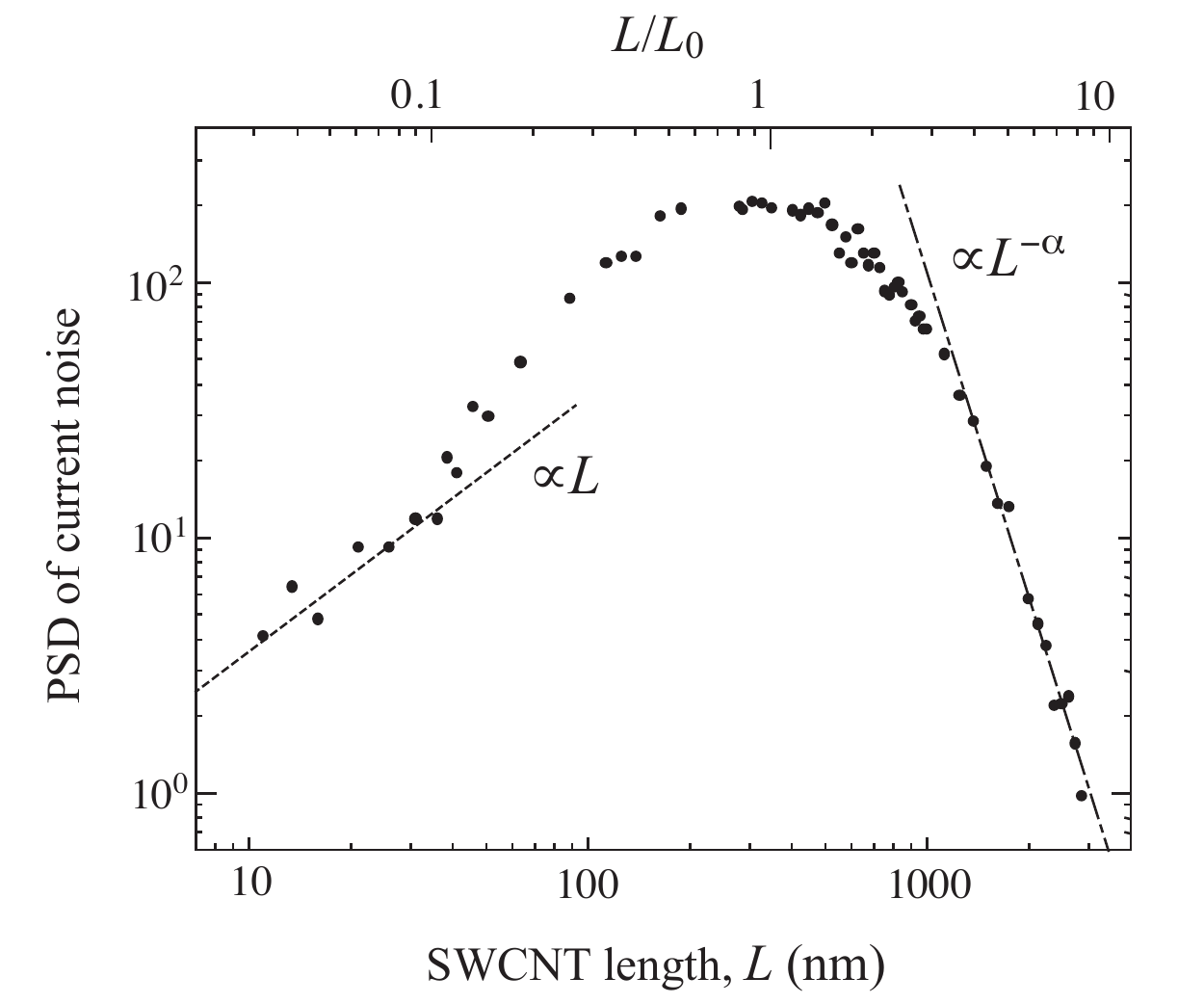}
\caption{Length dependence of ${\mathcal S}_{\rm KK}$ (dimensionless PSD of current noise) in (5,5) SWCNT at $300$~K.
The scaling parameter is numerically estimated to be $\alpha=3.81$. Here, $L_0=346$~nm is the mean free path of an electron. }
\label{fig:LD_fluc}
\end{center}
\end{figure}

By calculating the variance of cumulative charge $\langle\Delta q(\tau)\rangle=\langle q(\tau)^2\rangle-\langle q(\tau)\rangle^2$ at $t=\tau$, we can obtain 
the dimensionless PSD of the current noise, which is given by
\begin{equation}
{\mathcal S}_{\rm KK}=\frac{2\langle (\Delta q(\tau)^2)\rangle}{N_\tau},
\label{eq:s}
\end{equation}
where $N_\tau=\tau/\Delta t$ is the number of time steps within the interval $0\le t\le \tau$. We confirmed that ${\mathcal S}_{\rm KK}$ does not depend on $\tau$
for sufficiently large $\tau$, as in the case of $\tau=5.0$~ps used in this simulation.
Figure~\ref{fig:LD_fluc} shows the $L$ dependence of 
${\mathcal S}_{\rm KK}$ in (5,5) SWCNTs at $300$~K. Unlike the conductance in Fig.~\ref{fig:LD_conductance}, ${\mathcal S}_{\rm KK}$ does not vary 
monotonically with respect to $L$; it has a peak when the system length is comparable to $L_0$. The appearance of a peak at $L/L_0\approx 1$ is physically 
natural because, at this length, the frequency at which electrons transmit through an SWCNT without scattering becomes comparable to the frequency at which 
they are reflected, resulting in maximum uncertainty in the instantaneous current. In the quantum ballistic regime ($L/L_0\ll 1$), the PSD increases in proportion 
to $L$, as shown by the dashed line in Fig.~\ref{fig:LD_fluc}. This can be understood as follows. In a short system, electrons are 
hardly scattered by phonons and therefore the PSD in Eq.~(\ref{eq:landaure_S}) can be approximately expressed as
\begin{equation}
{\mathcal S}_{\rm KK}\propto \zeta_{\rm K}(1-\zeta_{\rm K})\approx 1-\zeta_{\rm K}
\label{eq:S_T(1-T)}
\end{equation}
using the effective transmission function $\zeta_{\rm K}(\approx 1)$. As seen from Eq.~(\ref{eq:datta}), the transmission function is given by 
$\zeta_{\rm K}\approx 1-L/L_0$ in the limit of $L/L_0\ll 1$. The PSD eventually varies as ${\mathcal S}_{\rm KK}\propto L$.  Note that
a similar result has been obtained based on Boltzmann transport theory~\cite{rr:Gurevich}.
In addition, we can see in 
Fig.~\ref{fig:LD_fluc} that when the system size $L/L_0$ exceeds about 0.1, ${\mathcal S}_{\rm KK}$ deviates from linearity with respect to $L$
and increases in a super-linear manner. 
This means that electron--phonon scattering has become apparent. It is thus no longer appropriate to express the PSD in terms of $\zeta_{\rm K}$, 
as shown in Eq.~(\ref{eq:S_T(1-T)}).

On the other hand, in the classical diffusive transport regime ($L/L_0\gg 1$), electron--phonon scattering brings a nonequilibrium system closer to local 
thermal equilibrium; therefore, we can expect the current noise to be suppressed with increasing $L$. In fact, we found in Fig.~\ref{fig:LD_fluc}
that the current noise of SWCNTs decreases with $L$ and obeys the scaling law ${\mathcal S}_{\rm KK}\propto L^{-\alpha}$. The scaling parameter 
was determined to be $\alpha=3.81$. Let us compare this result with previous theoretical results.
For example, Shimizu and Ueda concluded that the current noise decreases rapidly with 
$L$ as $e^{-(L/L_0)^2}$~\cite{rr:Shimizu}. On the other hand, based on the fictitious probe method, Beenakker and B{\" u}ttiker pointed out that the noise 
decreases more slowly and is proportional to $L^{-2}$~\cite{rr:Beenakker}. In addition, Nagaev and Naveh {\it et al}. argue that the current noise scales as $L^{-2/5}$ 
based on a simple model with an energy-dependent relaxation time $\tau \propto \varepsilon^3$~\cite{rr:Nagaev1992,rr:Nagaev1995,rr:Naveh}. 
At the very least, our results for SWCNTs support the power-law decay and suggest that the scaling parameter $\alpha$ is not universal, but rather depends on microscopic material parameters such as the nature of scattering processes, the strength of electron--phonon interactions, and the phonon dispersion relation.
Further theoretical studies from a microscopic perspective are desirable to clarify these aspects, which we leave for future work.

\section{Summary}
We investigated the phonon-induced current noise in SWCNTs within a wide range of tube length $L$ (from nano- to micrometers) using the Open-TDSE+MD 
method. We found that the current noise exhibits a maximum when $L\sim L_0$. This is because the ratio of electrons transmitted through 
an SWCNT without scattering is comparable to that of electrons being scattered and reflected. The mean free path of SWCNTs is typically around several hundred 
nanometers at room temperature, which is about the same length as that of SWCNTs used in electronic devices. Therefore, when designing electronic devices, it is 
essential to control current noise. We also found that the current noise increases in proportion to $L$ in the ballistic regime ($L\ll L_0$), whereas the power-law 
decay is $L^{-\alpha}$ with scaling parameter $\alpha=3.81$ in the diffusive regime ($L\gg L_0$). Apart from the value of the scaling parameter, we expect that 
the $L$ dependence in both ballistic and diffusive transport limits is universal for 1D quantum wires, not limited to SWCNTs. In the future, the analysis of 
time-dependent current based on the Open-TDSE+MD method will be extended to full-counting statistics, including the estimation of higher-order cumulants 
such as the skewness and kurtosis of the current histogram.

\section*{Acknowledgements}
We would like to thank Satofumi Souma, Kenji Sasaoka, Keisuke Ishizeki, and Raimu Akimoto for useful discussions regarding this research. We also thank 
Takeo Kato for his helpful comments on the draft of this paper. This work was supported in part by the Japan Society for the Promotion of Science 
KAKENHI (grant no. 23H00259) and by Toshiba Electronic Devices \& Storage Corporation under an Academic Encouragement Grant.

\end{document}